# Single photon Mach-Zehnder interferometer for quantum networks based on the Single Photon Faraday Effect: principle and applications

H P Seigneur[1], Michael N Leuenberger[2] and W V Schoenfeld[1]

[1]CREOL: College of Optics and Photonics, University of Central Florida
Orlando, FL, USA 32816
[2]Department of Physics, University of Central Florida Orlando, FL, USA 32816
E-mail: seigneur@creol.ucf.edu, mleuenbe@mail.ucf.edu and winston@creol.ucf.edu

**Abstract.** Combining the recent progress in semiconductor nanostructures along with the versatility of photonic crystals in confining and manipulating light, quantum networks allow for the prospect of an integrated and low power quantum technology. Within quantum networks, which consist of a system of waveguides and nanocavities with embedded quantum dots, it has been demonstrated in theory that many-qubit states stored in electron spins could be teleported from one quantum dot to another via a single photon using the Single Photon Faraday Effect. However, in addition to being able to transfer quantum information from one location to another, quantum networks need added functionality such as (1) controlling the flow of the quantum information and (2) performing specific operations on qubits that can be easily integrated. In this paper, we show how in principle a single photon Mach-Zehnder interferometer, which uses the concept of the single photon Faraday Effect to manipulate the geometrical phase of a single photon, can be operated both as a switch to control the flow of quantum information inside the quantum network and as various single qubit quantum gates to perform operations on a single photon. Our proposed Mach-Zehnder interferometer can be fully integrated as part of a quantum network on a chip. Given that the X gate, the Z gate, and the XZ gate are essential for the implementation of quantum teleportation, we show explicitly their implementation by means of our proposed single photon Mach-Zehnder interferometer. We also show explicitly the implementation of the Hadamard gate and the single-qubit phase gate, which are needed to complete the universal set of quantum gates for integrated quantum computing in a quantum network.
03.67.–a / 42.50.St / 42.82.Gw / 42.50.Pq / 71.70.Ej

# 1. Introduction

Single photon interferometry has been, and continues to be, a valuable method to study and understand a vast range of phenomena in physics, particularly in quantum mechanics such as *quantum superposition* (Wheeler and Zurek 1983), the "*quantum eraser*" which is the possibility of choosing the determination of paths after the interferometer (Scully and Druhl 1982), *dephasing* due to the quantum channel resulting from the inevitable interaction between the channel and the quanta of light (Xiang et al. 2006), or even the *quantum Sagnac Effect* (Bertocchi et al. 2006).

In addition, single photon interferometers can also play a key role as core components of new devices in the field of quantum information in a similar way that integrated Mach-Zehnder interferometers (MZ) are fundamental elements of photonic switches in optical communication. In fact, Shimizu and Imoto already proposed a single-photon interferometer for cryptographic quantum communication (Shimizu and Imoto 2000). Rather than manipulating and measuring the internal states of entangled photon twins, cryptographic quantum communication is possible by manipulating a single photon in an extended interferometer. Knill, LaFlamme and Milburn showed that efficient quantum computation is possible using only linear optical components (beam splitters and phase shifters -i.e. Elements that make up MZ interferometers), single photon sources, and photo-detectors (Knill et al. 2001).

Although the use of single photon interferometers has been verified in processing quantum information in the case of quantum key distribution (Shimizu and Imoto 2000) and, in principle, in quantum computation (Knill et al. 2001), they have been typically implemented in a way that was relatively *bulky* and *inadequate* for use within a high density integration setting. Yet, there is much interest in a quantum technology that can be implemented on a chip. One important reason is that an integrated quantum technology could in theory continue to fulfill Moore's law just by adding qubits to the Hilbert space (Nielsen and Chuang 2000), and as a result, constitute itself a potentially attractive solution to the current difficult challenges of further decreasing transistors size in order to fulfill Moore's law, that is assuming the cost of fabricating such quantum technology can be kept reasonable.

The approach for a quantum technology that we examine in this publication is based on quantum networks (Englund et al. 2006), which is deemed very promising given recent progresses in the conception and fabrication of semiconductor nanostructures and photonic crystals, along with single photon on-demand sources. One important implementation of quantum networks permits the teleportation of quantum information from qubits made of the spin of an excess electron in the conduction band of *quantum dots* (QD) embedded in nanocavities that interact via single photons by means of the *Single Photon Faraday Effect* (SPFE) (Leuenberger et al. 2005). Such spin based implementation of a quantum network allows for the prospect of an *integrated* and *low power* (Wolf et al. 2001) quantum technology in which *single-photon Mach-Zehnder interferometers* (SMZ) are envisioned to play a fundamental role.

In this paper, the operating principle and properties of SMZ interferometers and their applications are introduced. The operation of SMZ interferometers is based on the Single Photon Faraday Effect, presented in **Section 2**. By means of this effect, SMZ interferometers are able to manipulate the geometrical phase of a single photon as a *switching* mechanism in order to control the flow of quantum information within a quantum network. Using the geometrical phase instead of an optical path difference within a MZ interferometer is a bit unusual, however it allows for an additional functionality – it can also be used to create a superposition of the photon polarization eigenbasis or simply a change of eigenbasis (i.e. linear to circularly polarized light). In other words, SMZ interferometers could also be used as *single qubit gates*, and this is of critical importance for quantum networks in which single photons and their polarization eigenbasis are used to realize "messenger" qubits. These two separate applications of SMZ interferometers, namely an optical switch and a single qubit gate, are discussed in **Section 3**. Design issues, dynamic behavior, as well as the anticipated performance of SMZ interferometers are presented in **Section 4**. We conclude in **Section 5** by comparing the properties of the SMZ interferometer with existing implementations of switches and single qubit gates.

## 2. Theory

The *classical Faraday Effect* is a linear magneto-optic effect, which is characterized by a rotation of the linear polarization of light propagating inside an isotropic medium subject to an external constant magnetic field applied in the direction of propagation. The general concept behind the Faraday Effect is that a linearly polarized wave can be decomposed into two circularly polarized waves, which are the appropriate normal modes in this regime; each circularly polarized normal mode propagates with different refractive index. Quantum mechanics tells us that the magnetic field induced splitting of the energy levels with different total angular momentum is the reason for the linear polarization rotation, and hence the different circular polarizations of light couple differently during the process of virtual absorption, which is responsible for the existence of refractive indices. Similarly, the SPFE involves the rotation of linearly polarized light as a result of broken symmetry between the left and right components of circularly polarized light. However, the SPFE only involves the nonresonant interaction of a single photon with a two level system, does not require an external magnetic field, and can result in an entanglement of the photon with an electron spin (Leuenberger et al. 2005).

The two-level system of interest in **Figure 1** and **Figure 2** can be deduced under certain assumptions. First, the split-off band is purposely ignored since typical split-off energies are on the order of several hundreds of meV, thus bringing the energy level well out of resonance with the single photon. Second, spherical dots are assumed in order to maintain a *2-fold* energy degeneracy in the conduction band and a *4-fold* energy degeneracy in the valence band at the Γ point. Also, due to quantum size effects within the quantum dot, the effective bandgap is shifted to higher energies. Third, under the appropriate extrinsic doping and thermal conditions, it can be assumed that the top of the valence band is filled with four electrons while there is an excess electron in the conduction band. It is the presence of this extra conduction band electron that enables the SPFE. If the energy of the single photon is taken to be slightly below the effective band gap energy (i.e. slightly detuned), considering the parity condition imposed from the matrix element of envelop functions in semiconductor nanostructures, the transition from the top of the valence band (m=1) to the bottom of the conduction band (n=1) is the strongest transition by far. Since the matrix elements of other transitions are much weaker, one can then consider this transition as a two-level system.

We now isolate our consideration to this two-level system. Recalling that the single photon energy is slightly detuned from the two-level system resonance, the SPFE is a virtual process resulting from the transition rules that govern the *electric-dipole interaction*. In fact, it is the transition rules that allow the symmetry of circular polarized light to be broken as it couples to the two-level system during propagation, resulting in a geometrical phase shift of the single photon polarization. This is much different than the shift in energy level due to an external magnetic field exploited in the classical case. The ensuing geometrical phase shift is a consequence of one circular polarization, say right-hand circular, $\sigma^+$, interacting with only the heavy holes band and the other circular polarization, left-hand circular, $\sigma^-$, with the light holes band. Since the matrix element involving the heavy holes band is larger than the matrix element involving the light holes band, both circularly polarized components accumulate different phases resulting in a rotation of

the linear polarization. Using the Jaynes-Cummings model, the Hamiltonian of the system can be written as

$$\begin{aligned}
H = &\hbar\omega_{cav}\left(a^{\dagger}_{\sigma^+}a_{\sigma^+} + a^{\dagger}_{\sigma^-}a_{\sigma^-}\right) \\
&+ \hbar\omega_{hh}\sigma_{3/2v} + \hbar\omega_{hh}\sigma_{-3/2v} + \hbar\omega_{lh}\sigma_{1/2v} + \hbar\omega_{lh}\sigma_{-1/2v} + \hbar\omega_{e}\sigma_{1/2c} + \hbar\omega_{e}\sigma_{-1/2c} \\
&+ \hbar g_{3/2v,1/2c}\left(a^{\dagger}_{\sigma^-}\sigma_{3/2v,1/2c} + a_{\sigma^-}\sigma_{1/2c,3/2v} + a^{\dagger}_{\sigma^+}\sigma_{-3/2v,-1/2c} + a_{\sigma^+}\sigma_{-1/2c,-3/2v}\right) \\
&+ \hbar g_{1/2v,1/2c}\left(a^{\dagger}_{\sigma^-}\sigma_{1/2v,-1/2c} + a_{\sigma^-}\sigma_{-1/2c,1/2v} + a^{\dagger}_{\sigma^+}\sigma_{-1/2v,1/2c} + a_{\sigma^+}\sigma_{1/2c,-1/2v}\right)
\end{aligned} \quad (1)$$

where

$$\begin{aligned}
H_{Field} &= \hbar\omega_{cav}\left(a^{\dagger}_{\sigma^+}a_{\sigma^+} + a^{\dagger}_{\sigma^-}a_{\sigma^-}\right) \\
H_{Atom} &= \hbar\omega_{hh1,2}\sigma_{3/2v} + \hbar\omega_{hh1,2}\sigma_{-3/2v} + \hbar\omega_{lh1,2}\sigma_{1/2v} + \hbar\omega_{lh1,2}\sigma_{-1/2v} + \hbar\omega_{e1,2}\sigma_{1/2c} + \hbar\omega_{e1,2}\sigma_{-1/2c} \\
H_{J-C} &= \hbar g_{3/2v,1/2c}\left(a^{\dagger}_{\sigma^-}\sigma_{3/2v,1/2c} + a_{\sigma^-}\sigma_{1/2c,3/2v} + a^{\dagger}_{\sigma^+}\sigma_{-3/2v,-1/2c} + a_{\sigma^+}\sigma_{-1/2c,-3/2v}\right) \\
&+ \hbar g_{1/2v,1/2c}\left(a^{\dagger}_{\sigma^-}\sigma_{1/2v,-1/2c} + a_{\sigma^-}\sigma_{-1/2c,1/2v} + a^{\dagger}_{\sigma^+}\sigma_{-1/2v,1/2c} + a_{\sigma^+}\sigma_{1/2c,-1/2v}\right)
\end{aligned}$$

**Figure 1** and **Figure 2** show how different circularly polarized light interacts with the two-level system. The direction of the single-photon Faraday rotation is conditional on the spin orientation of the excess electron, i.e. it is clockwise if the spin is up (↑), and counter clockwise if the spin is down (↓). The single-photon Faraday rotation is a result of the different phase accumulated for the right circularly polarized RCP ($\sigma^+$) and the left circularly polarized LCP ($\sigma^-$) light during interaction with the two-level system or quantum dot (Leuenberger et al. 2005).

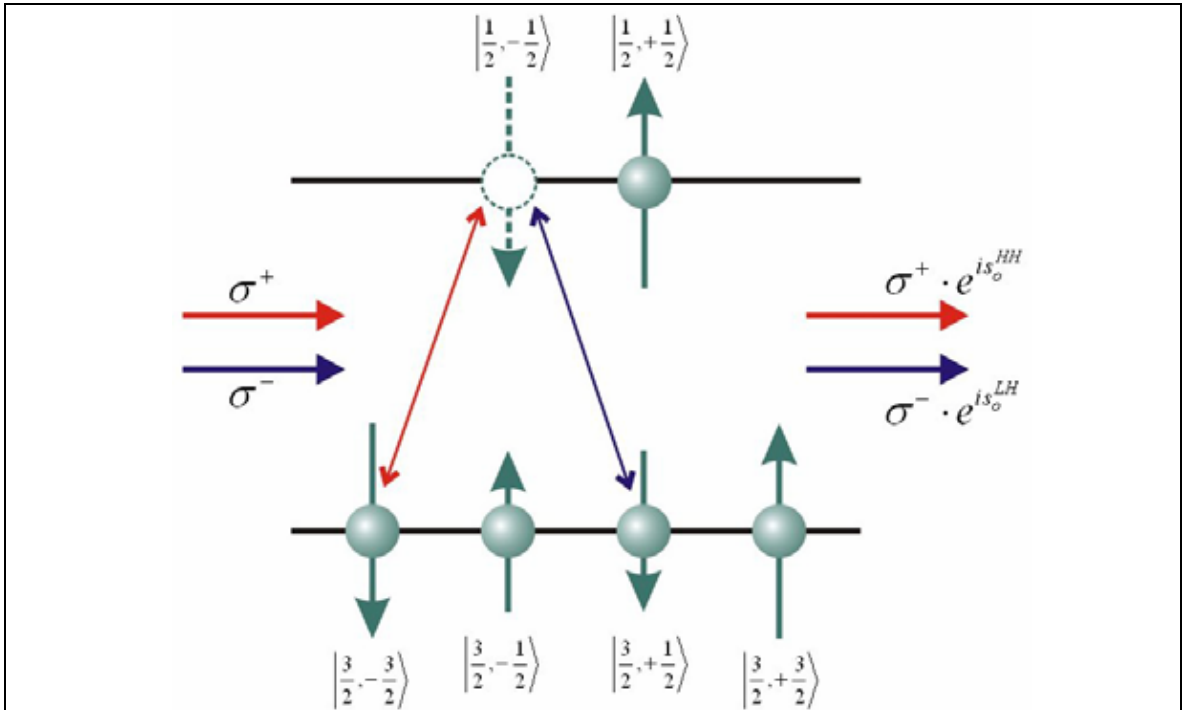

**Figure 1.**
The rotation of the single photon linear polarization is clockwise in the plane perpendicular to its propagation when the excess electron spin is up (↑).

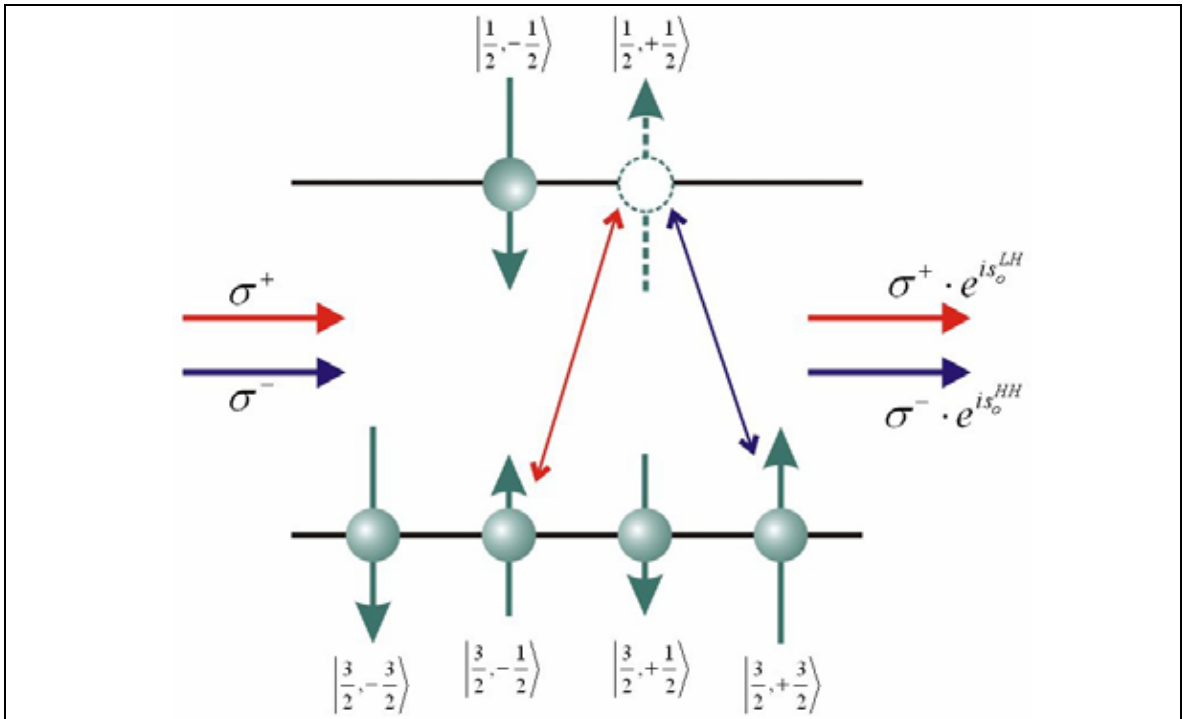

**Figure 2.**
The rotation of the single photon linear polarization is counterclockwise in the plane perpendicular to its propagation when the excess electron spin is down ($\downarrow$).

## 3. Applications

Interferometers usually are composed of two basic elements: (1) *passive structural components* for splitting and recombining the light and (2) *active components* for manipulating the phase of light. First, passive structural components for splitting and recombining the light considered in this paper are based on photonic crystals (see **Section 4** for details on design issues). In the case of the switch, the intensity of the photon is split in half in each arm; where as, in the case of the single qubit gate, the photon is split according to its polarization, possibly resulting in an unbalance in each arm in term of the intensity of the light. Second, the active components in a SMZ interferometer consist of a pair of nanocavities, each coupled to one arm of the SMZ interferometer. Each cavity and its embedded quantum dot provide the environment for the SPFE to take place and therefore provide the means for the manipulation of the photon geometrical phase. Upon completion of the desired amount of rotation of the linear polarization inside the nanocavities, the photon is released into the respective arms of the SMZ interferometer.

### 3.1. SMZ interferometer as a switch

**Figure 3** shows conceptually how the SMZ interferometer can be used as a switch. Contrary to the approach used in Ref (Leuenberger et al. 2005) in which the excess electron spin in the quantum dots within nanocavities are initialized to be in a superposition of up/down states for the purpose of entangling it with the single photon; here the spin is either up or down (in a pure state). There exists a variety of optical, electrical and magnetic techniques that can be used to initialize spins in semiconductors (Young et al. 2002). As for the quantization axis of the spins, it must be along the direction of propagation of the single photon.

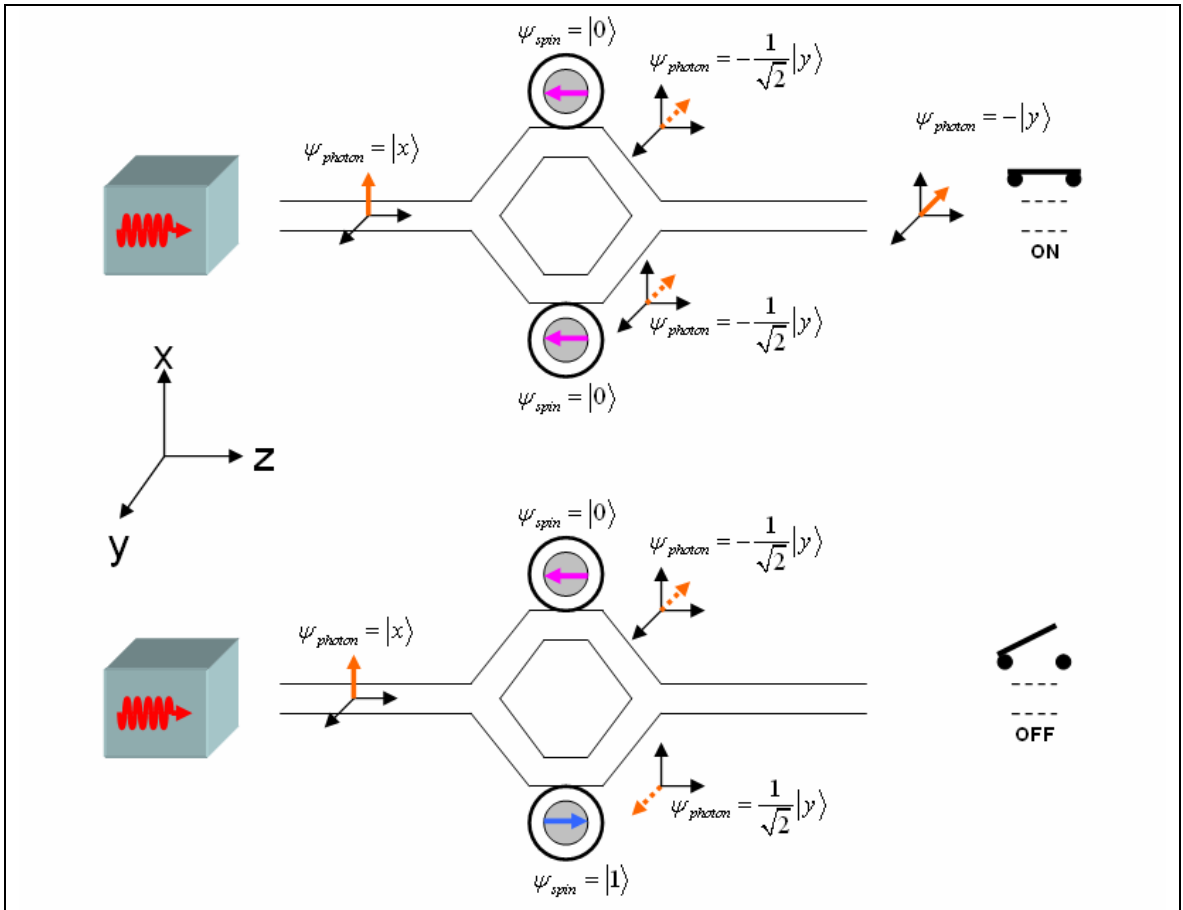

**Figure 3.**
Schematic for a spin-based SMZ interferometer where the spin of the electrons in the upper and lower arms are either parallel (above) or anti-parallel (below), giving respectively an identical and an opposite rotation of the polarization.

In its ON configuration, an initially x polarized (out of plane) photon is split into both arms of the SMZ interferometer, then in each arm the single photon couples inside nanocavities and interacts by means of the SPFE (strong coupling regime) with quantum dots that have their excess electron spin initialized in the down position or in –z direction (in plane). This causes its polarization to be rotated in each arm by $-90°$ pointing in the –y direction (in plane) after a specified interaction time, at which point the photon is released back into the waveguide (weak coupling regime). The photon is then able to interfere constructively and has its electric field positioned in the –y direction. A second possible ON configuration would involve the excess electron in the quantum dots having their spin parallel in the +z direction (or up position) instead of the –z direction as previously described resulting also in constructive interference but with the final polarization state of the photon in the +y direction. Conversely, the OFF configuration requires the spins to be antiparallel, one spin being initialized in the down position (–z direction) and the other in the up position (+z direction). The polarization in the arm with the spin down ($\downarrow$) is rotated by $-90°$ and pointing in the –y direction after a specified interaction time, while the polarization in the arm with the spin up ($\uparrow$) is rotated by $+90°$ and pointing in the +y direction. The photon is then able to interfere destructively at the output of the SMZ interferometer.

We now consider the scheme in **Figure 4** that depicts how an array of SMZ interferometers can be used to control the transfer of quantum information in parallel waveguides. Here the paths within the quantum network that have their SMZ interferometer switched OFF block the photon qubit from passing through while those that have their SMZ interferometer switched ON enable the transfer of quantum information by means of *GHZ quantum teleportation* (spin-photon-spin entanglement) (Leuenberger et al. 2005).

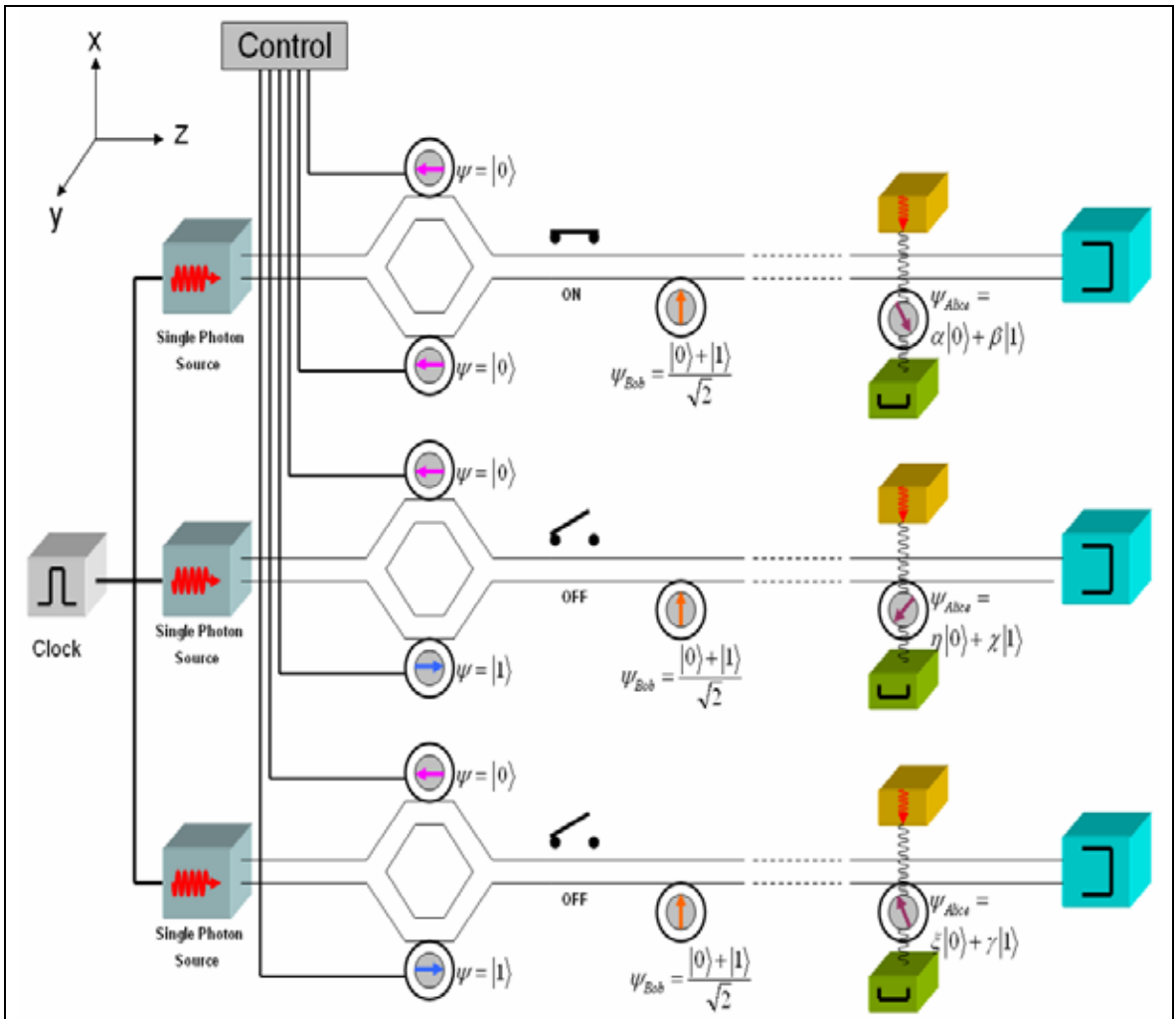

**Figure 4.**
The SMZ interferometer could be used as a switch. The top switch is ON and thus allows for the teleportation of Alice's qubit to Bob's qubit by means of the single photon produced at let say clock time t. However, the other switches are shown to be OFF and therefore the single photons produced from the sources aligned with these switches at the same time t are unable to pass through and carry on the transfer of quantum information.

Similarly, the SMZ interferometers can be used to achieve wavelength division multiplexing in quantum communication or quantum teleportation as shown in **figure 5**. In this case, the single photon sources can be designed such that each photon qubit they emit has a different wavelength and therefore can share the same channel or fiber. This requires that each SMZ interferometer has its quantum dot size and nanocavity size tuned such that its Faraday rotation rate and therefore its required interaction time remains the same as all the other interferometers independent of the wavelength at which they are being operated. This allows for a standardized technology in which all the components of the system can be synchronized.

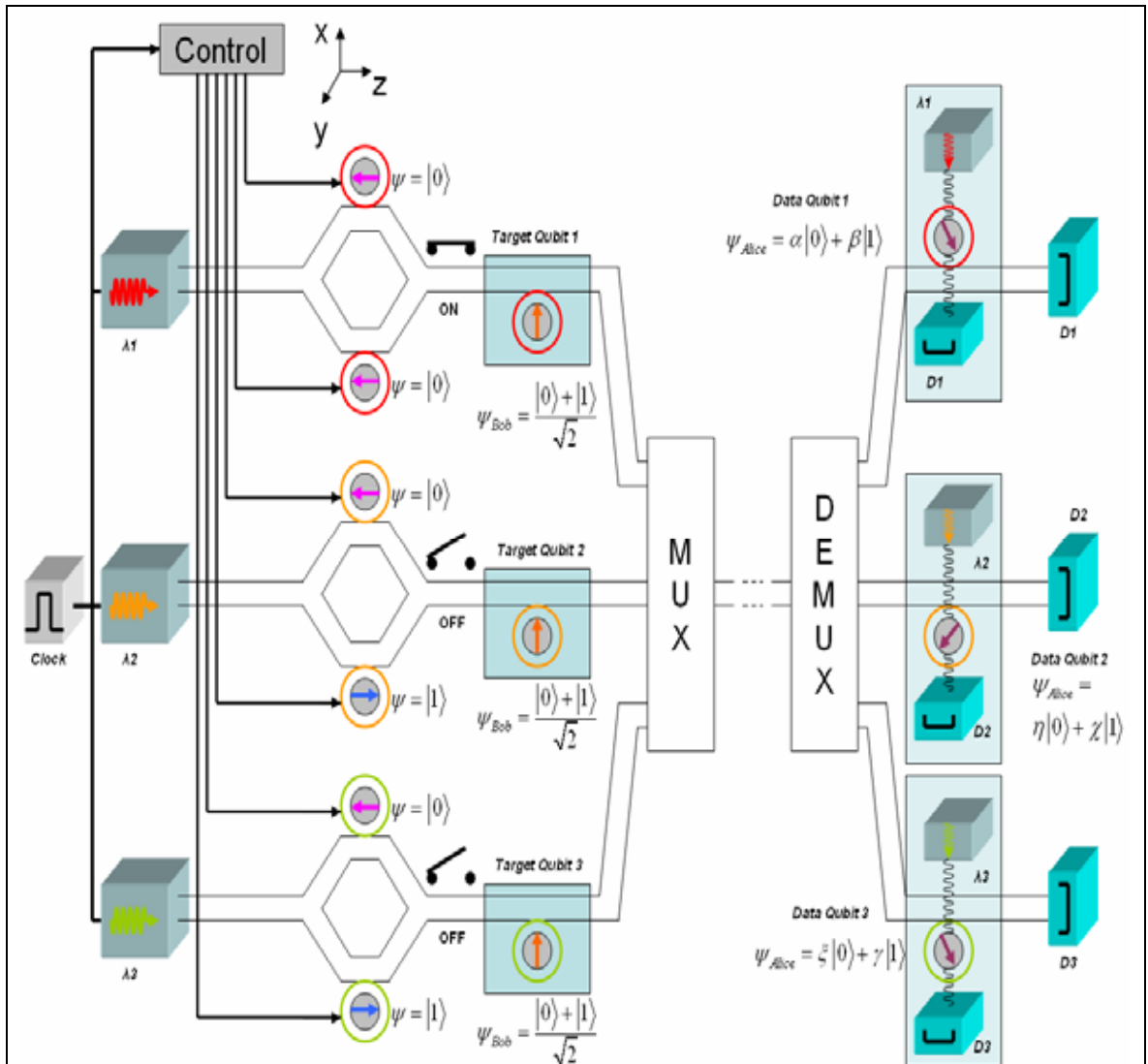

**Figure 5.**
Switches based on the SMZ interferometers could also be implemented in a Wavelength Division Multiplexing scheme for long distance quantum communication applications. Each single photon source is tuned to a different wavelength; then they are entangled with their respective Bob's qubit before the multiplexing step. At the destination node, the single photons are demultiplexed before they are allowed to interact with their respective Alice's qubit to complete the teleportation process.

*3.2. SMZ interferometer as a single qubit gate*
Furthermore, the SMZ interferometer can be used as a single qubit gate for the photon or "messenger" qubit. In quantum networks, information is stored in the spin of the excess electron in quantum dots located at specific sites. Keeping track of sites using any physical addressing scheme, this information can be accessed and retrieved when needed. This information can then be transfer to another location (or quantum dot) by means of a single photon that interacts with both the spins of the excess electrons at the quantum dots of origin and of destination creating a three particle entanglement or GHZ state resulting in teleportation (Leuenberger et al. 2005).

Single qubit gates, as their name indicates, consist of operations involving only one qubit. Mathematically, a quantum gate acting on a single qubit can be described by a 2 by 2 matrix, whereas the quantum state of the single qubit can be represented as a column vector. Because of the normalization condition that requires $|\alpha|^2 + |\beta|^2 = 1$ for the following quantum state $\alpha|0\rangle + \beta|1\rangle$, any matrix U corresponding to a single qubit gate must be unitary, that is $U^\dagger U = I$.

$$U\psi = \psi' \rightarrow U\begin{bmatrix}\alpha\\\beta\end{bmatrix} = \begin{bmatrix}\alpha'\\\beta'\end{bmatrix} \quad (2)$$

Classically, there is only one non-trivial member of this class called the NOT gate, which transforms a 0 into a 1 and a 1 into a 0. Similarly, a quantum gate that would transform the state $|0\rangle$ into the state $|1\rangle$, and vice versa could be considered a quantum NOT gate. However, in quantum information processing, we are interested in states that are in a superposition of eigenbasis. Due to this fact, unlike the classical case, there are technically an infinite number of non-trivial single qubit gates. Below, four important single qubit gates known as the X gate, the Z gate, the XZ gate, the Hadamard or H gate, and the Phase gate are considered. The X, Z, and XZ gates are needed for the reconstruction of a state in the teleportation protocol (Bennett et al. 1993). Note that our implementations of the X, Z, and XZ gate by means of the SMZ interferometer have not been shown before (Leuenberger et al. 2005). Our implementations are essential for the realization of a teleportation device that works only by means of the SPFE. Only then the teleportation method based on the SPFE can be completely integrated.

$$\alpha|0\rangle + \beta|1\rangle \xrightarrow{X} \beta|0\rangle + \alpha|1\rangle$$

$$\alpha|0\rangle + \beta|1\rangle \xrightarrow{Z} \alpha|0\rangle - \beta|1\rangle$$

$$\alpha|0\rangle + \beta|1\rangle \xrightarrow{XZ} \beta|0\rangle - \alpha|1\rangle \quad (3)$$

$$\alpha|0\rangle + \beta|1\rangle \xrightarrow{H} \frac{\alpha+\beta}{\sqrt{2}}|0\rangle + \frac{\alpha-\beta}{\sqrt{2}}|1\rangle$$

$$\alpha|0\rangle + \beta|1\rangle \xrightarrow{R(\phi)} \alpha|0\rangle + \beta \cdot e^{2\pi i\phi}|1\rangle$$

where

$$X = \begin{bmatrix}0 & 1\\1 & 0\end{bmatrix}, \quad Z = \begin{bmatrix}1 & 0\\0 & -1\end{bmatrix}, \quad XZ = \begin{bmatrix}0 & -1\\1 & 0\end{bmatrix},$$

$$H = \frac{X+Z}{\sqrt{2}} = \frac{1}{\sqrt{2}}\begin{bmatrix}1 & 1\\1 & -1\end{bmatrix}, \text{ and } R(\phi) = \begin{bmatrix}1 & 0\\0 & e^{2\pi i\phi}\end{bmatrix}$$

In order to realize single quantum gates, a polarizing beam splitter is needed within the SMZ interferometer as opposed to the initially considered beam splitter that split the beam in two halves of equal amplitudes. Until a few years ago, most of the approaches in realizing polarizing beam splitters typically required relatively large size structures (length of the order of millimeters), which was undesirable for an integrated quantum technology. However, Kim et al have proposed an ultra compact high-efficiency polarizing beam splitter that operates over a wide wavelength range and that is based on a hybrid photonic crystal and a conventional waveguide structure (Kim et al. 2003). Such technology could, without doubt, be implemented with the SMZ interferometer.

**Figure 6** considers the X gate. We identify the logic qubits by $|0\rangle = |x\rangle$, $|1\rangle = |y\rangle$. Given the following general initial state for the photon qubit $\psi_{photon} = \alpha|x\rangle + \beta|y\rangle$, the polarizing beam splitter split the incoming photon into two components; $\alpha|x\rangle$ goes into the lower arm while $\beta|y\rangle$ goes into the upper arm. Next, each polarization of the photon is rotated by 90° in opposite direction so that it corresponds to the other polarization eigenbasis (only two polarization eigenbasis in the given basis $|x\rangle$ *and* $|y\rangle$) and yet retains the same amplitude. In other words, if the first quadrant is considered (see **Figure 7**), then the y component of the polarization of the photon is rotated clockwise to $|y\rangle$ while the y component is rotated counterclockwise to $|x\rangle$. After the different components of the photon polarization are recombined, the final state of the photon is $\psi_{photon} = \beta|0\rangle + \alpha|1\rangle$. In respect to timing, the X gate takes less than 100 ps (see **Section 4**), which is approximately the interaction time needed to rotate both polarization components by $90°$. The four possible cases or quadrants for the X gate are shown in **Figure 7**, **Figure 8**, **Figure 9**, and **Figure 10**.

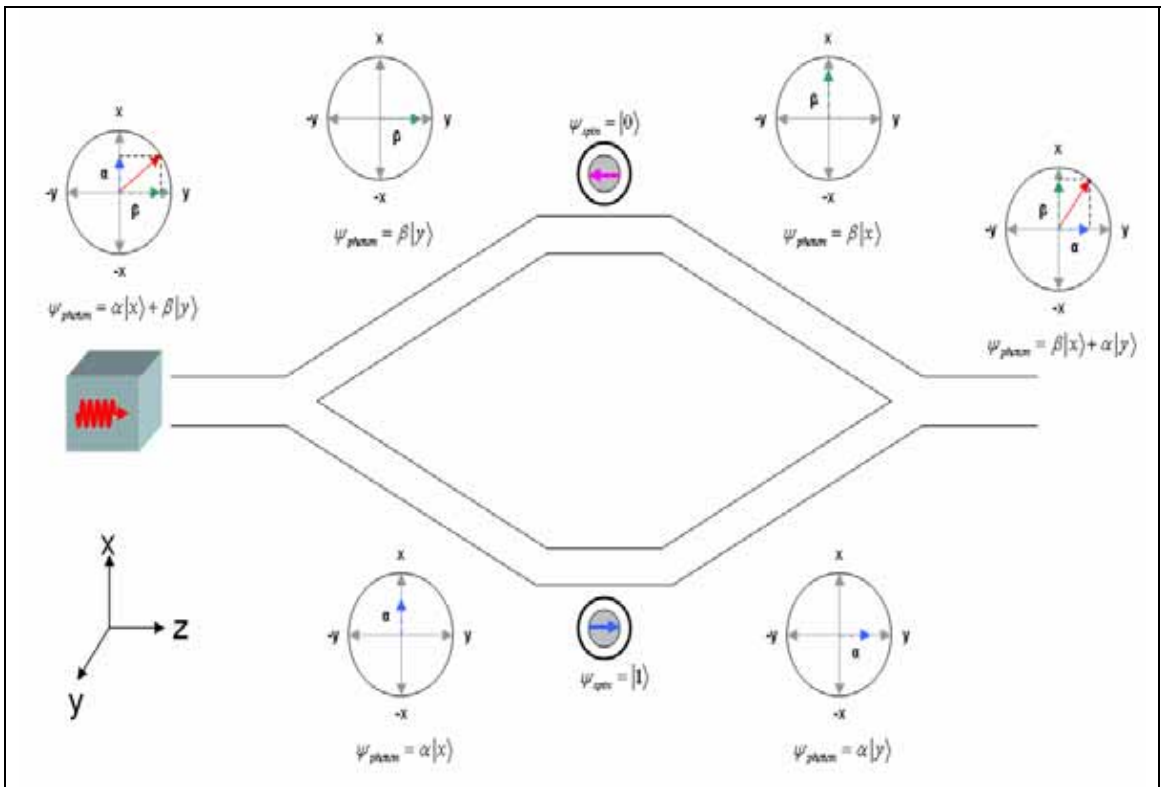

**Figure 6.**
SMZ interferometer as an X gate

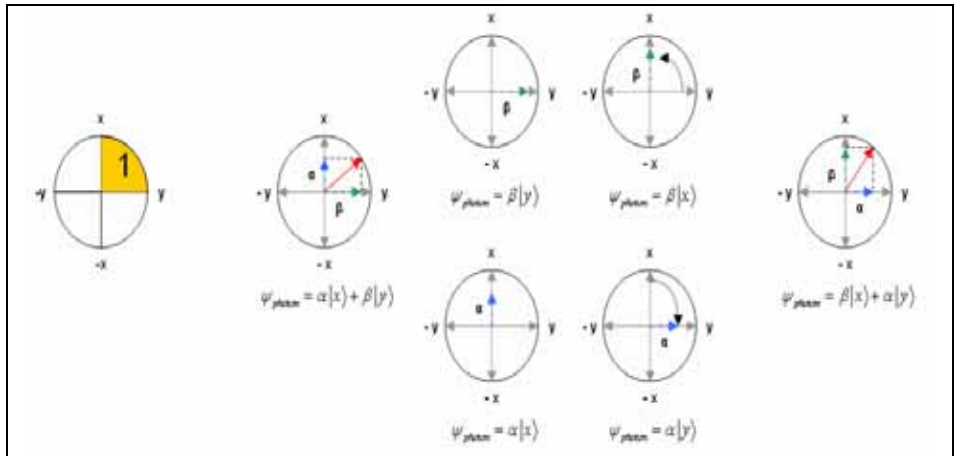

**Figure 7.**
SMZ interferometer as an X gate. The initial state of the single photon linear polarization is in the 1$^{st}$ quadrant.

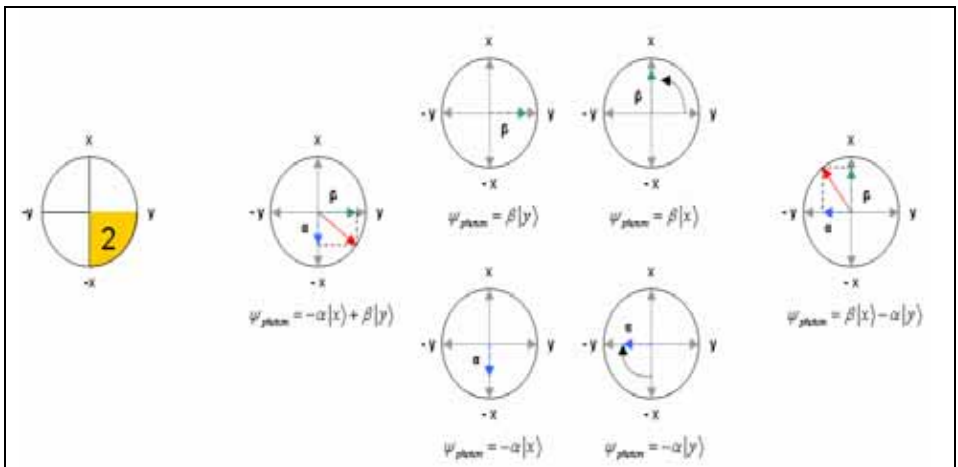

**Figure 8.**
SMZ interferometer as an X gate.
The initial state of the single photon linear polarization is in the 2$^{nd}$ quadrant.

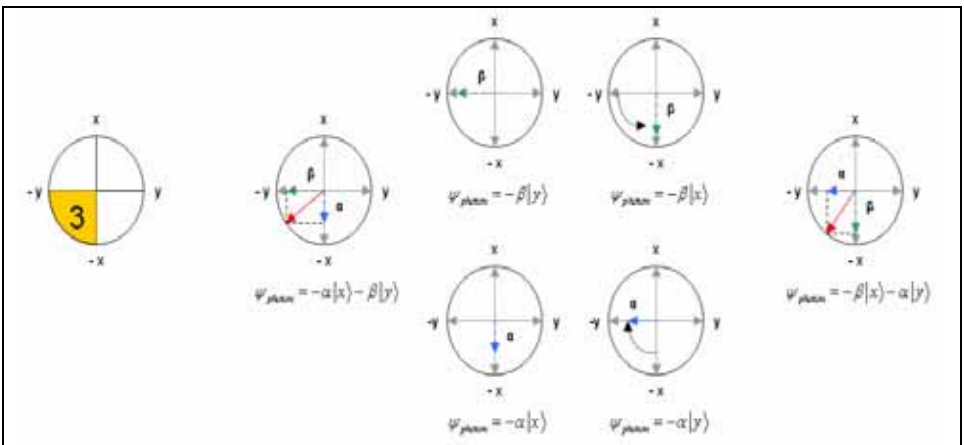

**Figure 9.**
SMZ interferometer as an X gate.
The initial state of the single photon linear polarization is in the 3$^{rd}$ quadrant.

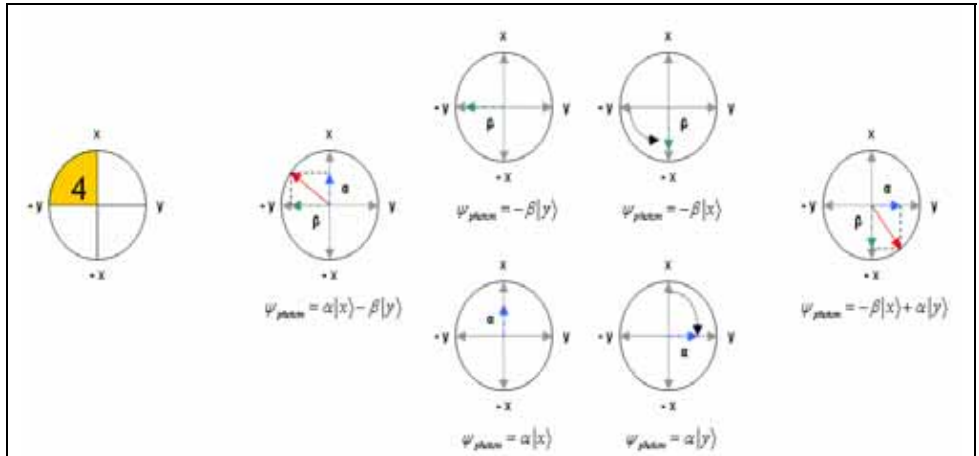

**Figure 10.**
SMZ interferometer as an X gate.
The initial state of the single photon linear polarization is in the 4$^{th}$ quadrant.

A schematic of the Z gate is shown in **Figure 11**. Here we consider again that the general initial state for the photon qubit is $\psi_{photon} = \alpha|x\rangle + \beta|y\rangle$ and that the polarizing beam splitter splits the incoming photon into two components, $\alpha|x\rangle$ and $\beta|y\rangle$. Next, in the case of the Z gate, only the polarization of the y component is rotated by 180°. After the different components of the photon polarization are recombined, the final state of the photon is $\psi_{photon} = \alpha|x\rangle - \beta|y\rangle$. Similarly, the Z gate takes less than 100 ps (see **Section 4**) to act on the photon qubit state.

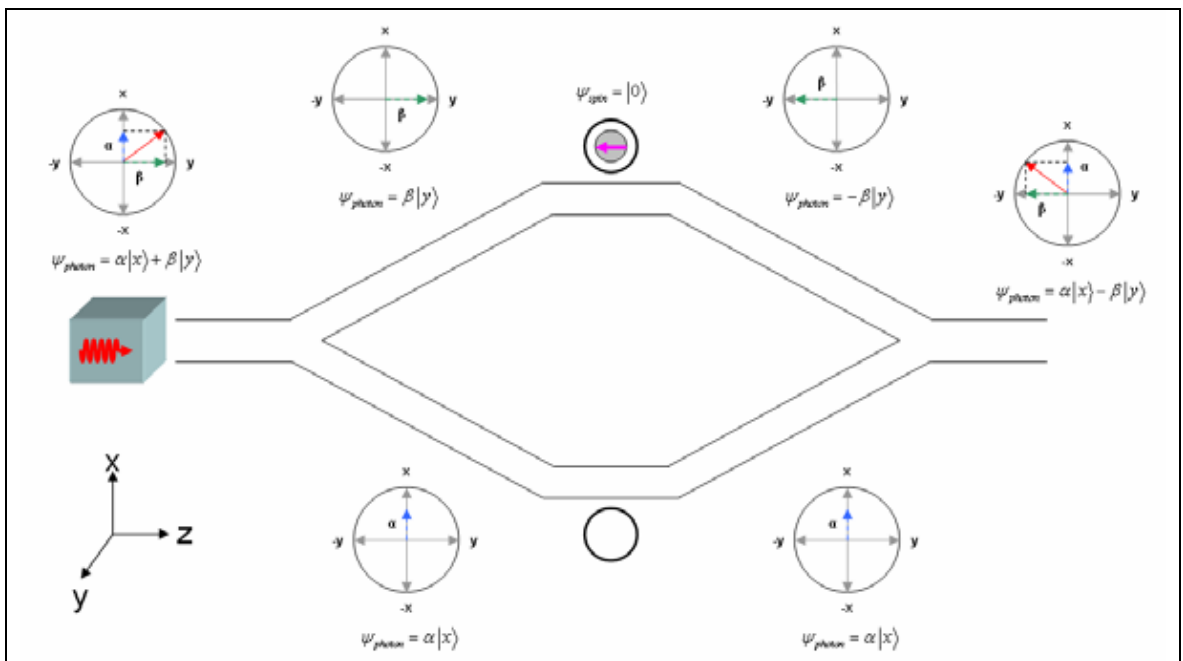

**Figure 11.**
SMZ interferometer as a Z gate.

A schematic of the XZ gate is shown in **Figure 12**. Here we consider again that the general initial state for the photon qubit is $\psi_{photon} = \alpha|x\rangle + \beta|y\rangle$ and that the polarizing beam splitter splits the incoming photon into two components, $\alpha|x\rangle$ and $\beta|y\rangle$. Next, in the case of the Z gate, each polarization component of the photon is rotated by 90° in the same counterclockwise direction (from the +y direction to the +x direction). After the different components of the photon polarization are recombined, the final state of the photon is $\psi_{photon} = \beta|x\rangle - \alpha|y\rangle$. Similarly, the XZ gate takes less than 100 ps (see **Section 4**) to act on the photon qubit state.

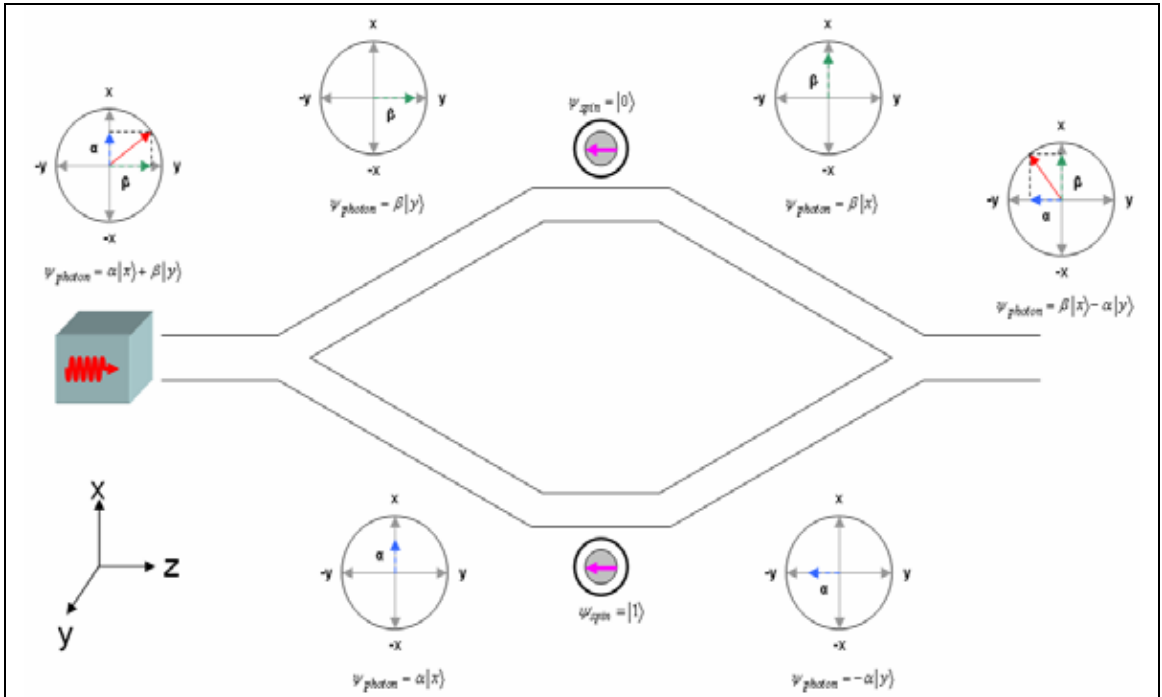

**Figure 12.**
SMZ interferometer as a XZ gate.

A schematic of the H gate is depicted in **Figure 13**. The photon qubit is again initialized to $\psi_{photon} = \alpha|x\rangle + \beta|y\rangle$, and the polarizing beam splitter splits the incoming photon into two components, $\alpha|x\rangle$ and $\beta|y\rangle$. Unlike the other gates, the H gate requires two stages. In the first stage, each polarization component of the photon is rotated by 90° in opposite direction. More precisely, the X component of the polarization of the photon is rotated clockwise while the Y component is rotated counterclockwise. In the second stage, each polarization component of the photon is rotated by only 45° in the same counterclockwise direction. After the different components of the photon polarization are recombined, the final state of the photon is $\psi_{photon} = \alpha \frac{|0\rangle + |1\rangle}{\sqrt{2}} + \beta \frac{|0\rangle - |1\rangle}{\sqrt{2}}$.

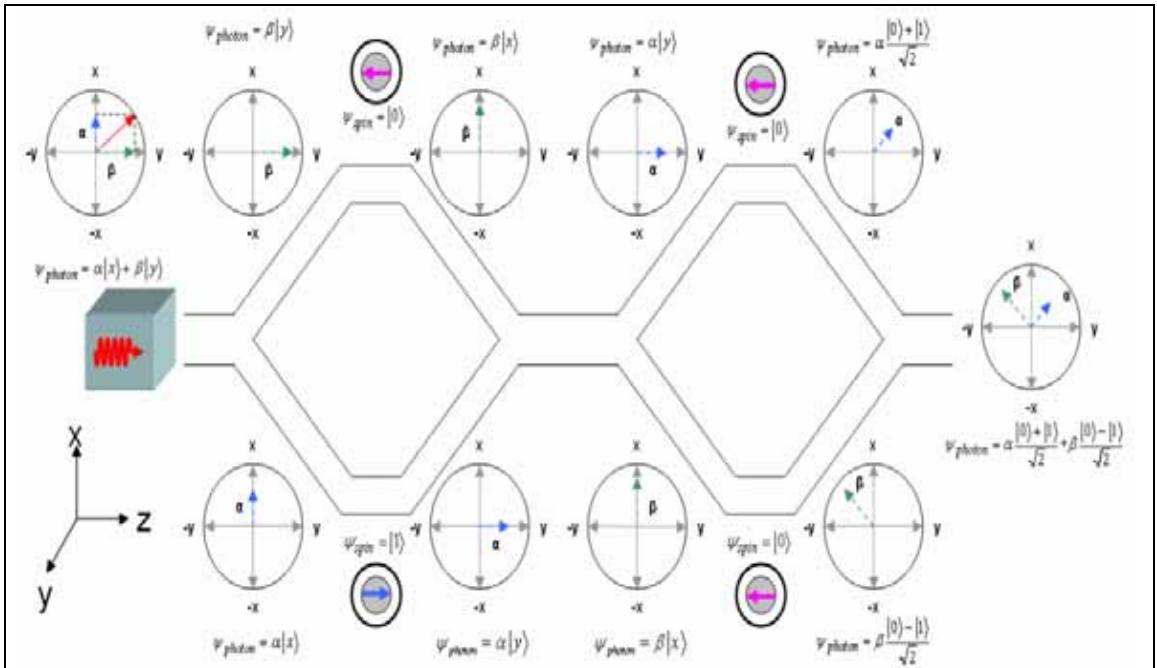

**Figure 13.**
SMZ interferometer as an H gate.

Last, the Phase gate is shown in Figure 14. The x and y polarizations are converted into right and left circular polarizations by means of $\frac{\lambda}{4}$ elements, which further splits the x and y polarizations in two equal magnitudes within their respective arm before inducing a $\frac{\pi}{2}$ phase resulting in LCP for the x polarization and RCP for the y polarization. The single spin leads then to a pure phase shift between the right and the left circular polarization, which can be converted back to x and y polarizations by means of more $\frac{\lambda}{4}$ elements. Also, the entire operation takes less than 100ps (see **Section 4**).

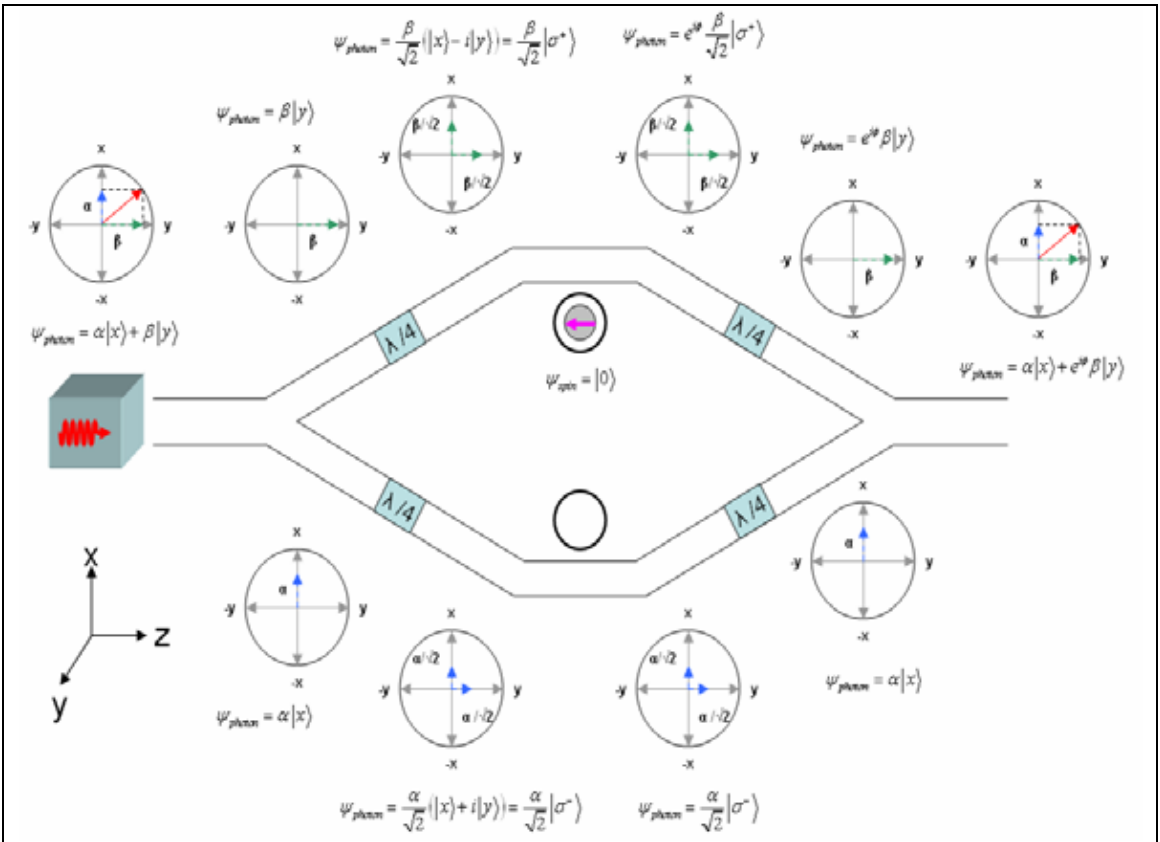

**Figure 14.**
SMZ interferometer as a Phase gate.

## 4. Design and Performance
### 4.1. Design
For a feasible realization of such switches and gates, a traveling photon propagating within a waveguide must easily couple in an adjacent nanocavity; in which it must remain trapped for a specific interaction time before it is released in a way that minimizes the phase error back into the waveguide. There potentially exist several approaches to implement the devices under investigation. One possible scheme utilizes extremely high Q cavities, which are assumed to be large enough to prevent the strong coupling regime. This *weak-coupling approach* to the SPFE is not desirable because of the large size of the cavities and the challenge of extremely high Q cavities in providing an environment that favors short interaction times. In this publication, an alternate approach that is more suitable for large scale integration is proposed. This method is attractive in a large scale integration setting because its aim is to employ smaller cavity sizes, faster rates for the rotation of the linear polarization, and smaller Qs. This method is called the *strong-coupling approach* to the SPFE, and the photon is sent in at one of the cavity polaritons.

The strong-coupling approach to the SPFE requires that a largely detuned photon strongly interacts with the quantum dot (2-level system) inside the cavity, resulting in its linear polarization undergoing a faster conditional rotation consistent with the spin state of the excess electron in the conduction band. Such interaction calls for the *strong coupling regime* (SCR), which requires that the coupling constant, $g$, between the quantum dot and the cavity mode to be much larger than both the dipole dephasing rate, $\gamma$, and the decay rate of the cavity, $\Gamma$ (i.e. $g \gg (\gamma, \Gamma)$). Furthermore, for the photon to be released, $g$ needs to be smaller than $\gamma$ and $\Gamma$, which is known as the *weak coupling regime* (WCR) (Khitrova et al. 2006). Naturally, for a given material system, cavity volume, and wavelength, both the coupling constant, $g$, and the dipole dephasing rate, $\gamma$, are set and cannot be changed during device operation in a controllable way. Only the cavity decay rate, $\Gamma$, is left in order to control the coupling regime in which the device is operating. Real-time lowering of the cavity decay rate (or increasing the cavity Q) promotes the SCR while increasing of the cavity decay rate (or lowering the cavity Q) allows for the WCR. Therefore, by actively or passively controlling the Q of the cavity, we can trap or release the photon at a given time.

For the physical realization of these SMZ based devices, the InAs/GaAs material system is considered. InAs self-assembled quantum dots are used to embody the two-level systems, which are embedded in high-Q GaAs photonic crystal nanocavities. Photonic crystals offer many potential advantages such as smaller cavity sizes (high electric field suitable for the SCR with a single QD) as well as strong confinement (minimizing in-plane losses) and ultra-compactness (sharp waveguide bend possible), thus providing the means for an integrated quantum technology. The ability to strongly confine light propagating in the plane within a 2D photonic crystal structure is based on the existence of a photonic bandgap due to a periodic change in the refractive index of the medium (optical modes whose frequencies are in the photonic bandgap are unable to propagate within the photonic crystal). Because the SPFE requires the rotation of the linear polarization of the photon in a plane perpendicular to the direction of propagation, the photonic crystal structure used to implement both the switch and the gates must support x (out of plane) and y (in plane) polarizations. As a result, the photonic crystal structure must consist of a triangular lattice of air columns in GaAs, which has overlapping of both TE (Electric field along y-axis or in-plane) and TM (Electric field along x-axis or out-of-plane) photonic bandgaps, as opposed to only one or the other for square lattices (Joannopoulos 1995). Furthermore, the optical mode of interest is assumed to be far enough away from the band edges of both the TE and TM bandgaps, since modes near band edges have smaller group velocities resulting in dramatic increase in Qs. For instance, modes closer to a TE band edges will bring about a higher Q value for TE modes with respect to TM modes. Also, TE modes are subject to more out-of-plane radiation losses than TM modes. In terms of phonic crystal cavity Qs defined in **Equation 4**, it means that the out-of-plane Q ($Q_\perp$) for TE mode is smaller than the out-of-plane Q for TM mode, while their in-plane Qs are the same.

$$Q = Q_\parallel + Q_\perp \qquad (4)$$

The out-of-plane radiation losses can be eliminated so that Qs for both TE and TM modes are matched by tailoring the mode profile using air-hole shifts and multi-step heterostructures (Asano et al. 2006). Finally, TE dipole modes of L1 cavities are degenerate in theory (The y-polarized mode and the z polarized mode), however because of imperfections due to fabrication these optical modes split into different eigenenergies. Nevertheless, the resulting splitting is very small and the Qs for each mode do not vary significantly and can be considered the same; therefore, as long as the photon propagates in the plane (y or z direction), the scheme remains unaffected (Shirane et al. 2007).

Before the cavity Qs and the cavity-waveguide coupling constant can be solved for, the dynamics of the 2-level system (QD) interacting with a single mode quantum field (photon) must be investigated and the rate of rotation of the linear polarization determined. The Hamiltonian given in Equation 1 describes how the RCP and LCP components of a linearly polarized photon field interact with the degenerate QD levels in a cavity. For simplicity, it is assumed that the spin of the excess electron in the conduction band is up ($\uparrow$). At any time t, the state vector $|\psi(t)\rangle$ is a linear combination of the states $|\uparrow,hh\rangle, |\uparrow,lh\rangle, |\uparrow,\sigma_z^+\rangle,$ and $|\uparrow,\sigma_z^-\rangle$. Here, $|\uparrow,hh\rangle$ and $|\uparrow,lh\rangle$ are the states

in which the QD is in an excited state with an heavy hole exciton (hh) or a light hole exciton (lh) and the spin of the excess electron in the conduction band is up. $|\uparrow,\sigma_z^+\rangle$ and $|\uparrow,\sigma_z^-\rangle$ are the states in which the QD dot is in the ground state with the photon present in the cavity, and the spin of the excess electron in the conduction band is up. The state vector is therefore

$$|\psi(t)\rangle = \frac{1}{\sqrt{2}}\left\{\left[C_{\uparrow hh}(t)|\uparrow,hh\rangle + C_{\uparrow\sigma^+}(t)|\uparrow,\sigma_z^+\rangle\right] + \left[C_{\uparrow lh}(t)|\uparrow,lh\rangle + C_{\uparrow\sigma^-}(t)|\uparrow,\sigma_z^-\rangle\right]\right\} \quad (5)$$

The equations of motion for the probability amplitudes $C_{\uparrow hh}(t), C_{\uparrow lh}(t), C_{\uparrow\sigma^+}(t),$ and $C_{\uparrow\sigma^-}(t)$ can easily derived and then solved exactly subject to certain initial conditions. The expressions for the probability amplitudes are

$$C_{\uparrow hh}(t) = \left\{\left[\cos(\frac{\Omega_{3/2}t}{2}) - \frac{i\Delta}{\Omega_{3/2}}\sin(\frac{\Omega_{3/2}t}{2})\right]C_{\uparrow hh}(0) - \left[\frac{i2g_{3/2v,1/2c}}{\Omega_{3/2}}\sin(\frac{\Omega_{3/2}t}{2})\right]C_{\uparrow\sigma^+}(0)\right\}e^{-i\cdot\frac{\Delta\cdot t}{2}}$$

$$C_{\uparrow\sigma^+}(t) = \left\{\left[\cos(\frac{\Omega_{3/2}t}{2}) + \frac{i\Delta}{\Omega_{3/2}}\sin(\frac{\Omega_{3/2}t}{2})\right]C_{\uparrow\sigma^+}(0) - \left[\frac{i2g_{3/2v,1/2c}}{\Omega_{3/2}}\sin(\frac{\Omega_{3/2}t}{2})\right]C_{\uparrow hh}(0)\right\}e^{-i\cdot\frac{\Delta\cdot t}{2}}$$

$$C_{\uparrow lh}(t) = \left\{\left[\cos(\frac{\Omega_{1/2}t}{2}) - \frac{i\Delta}{\Omega_{1/2}}\sin(\frac{\Omega_{1/2}t}{2})\right]C_{\uparrow lh}(0) - \left[\frac{i2g_{1/2v,1/2c}}{\Omega_{1/2}}\sin(\frac{\Omega_{1/2}t}{2})\right]C_{\uparrow\sigma^-}(0)\right\}e^{-i\cdot\frac{\Delta\cdot t}{2}} \quad (6)$$

$$C_{\uparrow\sigma^-}(t) = \left\{\left[\cos(\frac{\Omega_{1/2}t}{2}) + \frac{i\Delta}{\Omega_{1/2}}\sin(\frac{\Omega_{1/2}t}{2})\right]C_{\uparrow\sigma^-}(0) - \left[\frac{i2g_{1/2v,1/2c}}{\Omega_{1/2}}\sin(\frac{\Omega_{1/2}t}{2})\right]C_{\uparrow lh}(0)\right\}e^{-i\cdot\frac{\Delta\cdot t}{2}}$$

where

$\Omega_{3/2}^2 = \Delta^2 + 4\cdot g_{3/2v,1/2c}^2$ with $g_{3/2v,1/2c}$ being the coupling strength involving a heavy hole electron

$\Omega_{1/2}^2 = \Delta^2 + 4\cdot g_{1/2v,1/2c}^2$ with $g_{1/2v,1/2c}$ being the coupling strength involving a light hole electron

$g_{3/2v,1/2c} = 3\cdot g_{1/2v,1/2c}$

$\Delta = \omega - \omega_{ph}$ being the detuning frequency

The rate of rotation of the linear polarization is proportional to the difference phase accumulated for the RCP and LCP components of the linear polarization during interaction with the QD with the interaction frequencies $\Omega_{3/2}$ and $\Omega_{1/2}$, respectively. The coupling constants, $g$, between the QD and the photon field can easily be obtained for GaAs. In order to carry out this calculation, the preferred cavity for the realization of the switch and the gates is the L1 cavity (a single point defect). They do not necessarily possess Qs as high as L3 cavities (3 point defects along the $\Gamma-K$ direction), however they usually provide the smallest mode volume (Andreani et al. 2005). Shirane et al reported a single-defect or L1 nanocavities based on a triangular lattice GaAs photonic-crystal membrane with mode volume of $V = 0.039\mu m^3$ and a Q of 17,000 (Shirane et al. 2007). Additionally, InAs quantum dots less than 25 nm in size are reported to have dipole moment $\mu_{hh} = 29$ (in Debye) with bandgap energy $E_g = 1.32eV$, which correspond to an emission wavelength of $\lambda = 1.182\mu m$ (Khitrova et al. 2006).

If we consider the configuration of the switch in **Figure 3**, the single photon qubit is initially linearly polarized along the x axis or $\hat{e}_x = |x\rangle$. The expression for this linearly polarized photon can be rewritten in a different basis; one in which the linear polarization is considered as a superposition of right and left circular polarization:

$$\vec{e} = |x\rangle = \frac{1}{\sqrt{2}}\left[(\frac{|x\rangle + i|y\rangle}{\sqrt{2}}) + (\frac{|x\rangle - i|y\rangle}{\sqrt{2}})\right] = \frac{1}{\sqrt{2}}\left(|\sigma_z^+\rangle + |\sigma_z^-\rangle\right) \quad (7)$$

Once the photon interacts with the quantum dots in the separate arms of the SMZ interferometer, the right and left circular polarization components of the linear polarized photon accumulate different phases due to the single photon Faraday rotation effect. Assuming that the spin of excess electron in the conduction band of the quantum dot is up (↑):

$$\vec{e} = \frac{\left(e^{-iS_O^{HH}}|\sigma_z^+\rangle + e^{iS_O^{LH}}|\sigma_z^-\rangle\right)}{\sqrt{2}} \quad (8)$$

where $s_O^{HH}$ and $s_O^{LH}$ represent the phase shift for the photon when interacting with the heavy hole or light hole, respectively. It is useful to rewrite the phase in order to be able to derive an expression for the effective angle, $\varphi$, by which the linear polarization is rotated:

$$\vec{e} = e^{i\theta}\frac{\left(e^{-i\varphi}|\sigma_z^+\rangle + e^{i\varphi}|\sigma_z^-\rangle\right)}{\sqrt{2}} \quad (9)$$

where $\theta = \frac{(s_O^{HH} + s_O^{LH})}{2}$ and $\varphi = \frac{(s_O^{HH} - s_O^{LH})}{2}$.

In terms of the linear polarization basis eigenstates, the polarization can be expressed as:

$$\vec{e} = e^{i\theta}\left(\cos\varphi|x\rangle + \sin\varphi|y\rangle\right) \quad (10)$$

For the photon to destructively interfere with itself, the polarization initialized along the x axis must be rotated 90° in one arm so that it points in the y direction (spin of excess electron is ↑) and −90° in the other arm so that it points in the −y direction (spin of excess electron is ↓). This means that the angle $\varphi$ must be $\frac{\pi}{2}$ in one arm and $-\frac{\pi}{2}$ in the other arm. An expression for the phase shift accumulated for RCP and LCP components of the linearly polarized photon field during the interaction with the quantum dot in the nanocavity can be derived from **Equation 6**. In order to solve for the phase accumulated for the RCP component ($S_0^{HH}$), the time evolution of the probability $C_{\uparrow\sigma^+}(t)$ is used; whereas, the time evolution of the probability $C_{\uparrow\sigma^-}(t)$ is used for the phase accumulated for the LCP component ($S_0^{LH}$). Assuming the initial conditions are the following, $C_{\uparrow hh}(0) = 0$, $C_{\uparrow\sigma^+}(0) = 1$, $C_{\uparrow lh}(0) = 0$, and $C_{\uparrow\sigma^-}(0) = 1$, then the probability amplitudes of interest can be written as

$$C_{\uparrow\sigma^+}(t) = e^{-i\frac{\Delta \cdot t}{2}}\left[\cos(\frac{\Omega_{3/2}t}{2}) + \frac{i\Delta}{\Omega_{3/2}}\sin(\frac{\Omega_{3/2}t}{2})\right]C_{\uparrow\sigma^+}(0) \quad (11)$$

$$C_{\uparrow\sigma^-}(t) = e^{-i\frac{\Delta \cdot t}{2}}\left[\cos(\frac{\Omega_{1/2}t}{2}) + \frac{i\Delta}{\Omega_{1/2}}\sin(\frac{\Omega_{1/2}t}{2})\right]C_{\uparrow\sigma^-}(0)$$

Rewriting the complex coefficient $\left[\cos(\frac{\Omega_{3/2}t}{2}) + \frac{i\Delta}{\Omega_{3/2}}\sin(\frac{\Omega_{3/2}t}{2})\right]$ within the expression for the probability amplitude $C_{\uparrow\sigma^+}(t)$ in its exponential form using Euler's formula, an expression for the phase accumulated during the interaction of the RCP component with the heavy holes band can be obtained.

$$S_0^{HH} = \tan^{-1}\left(\frac{\Delta}{\Omega_{3/2}} \cdot \tan(\frac{\Omega_{3/2}t}{2})\right) \tag{12}$$

Similarly, an expression for the phase accumulated during the interaction of the LCP component with the light holes band can be obtained.

$$S_0^{LH} = \tan^{-1}\left(\frac{\Delta}{\Omega_{1/2}} \cdot \tan(\frac{\Omega_{1/2}t}{2})\right) \tag{13}$$

Since the relative phase $\varphi = \frac{(s_0^{HH} - s_0^{LH})}{2}$ must be $\pm\frac{\pi}{2}$, then $(s_0^{HH} - s_0^{LH})$ must be equal to $\pi$. Moreover, because we do not want to produce an exciton ensuing the completion of the rotation of the linear polarization rotation, it must be required that $\frac{\Omega_{3/2}T}{2} = j\pi$ and $\frac{\Omega_{1/2}T}{2} = j\pi$ in **Equation 6**. It was found that a detuning energy of $E_d = 75\mu eV$ gives an optimized operating point for both the phase requirement- $\varphi = \pi$ (see **Figure 15**) and the probability amplitude requirements- $\frac{\Omega_{3/2}T}{2} = j\pi$ and $\frac{\Omega_{1/2}T}{2} = j\pi$ (see **Figure 16**), resulting in an interaction time $T = 43 ps$. Consequently, it can be said that the *switching speed* is on the order of tens of picoseconds.

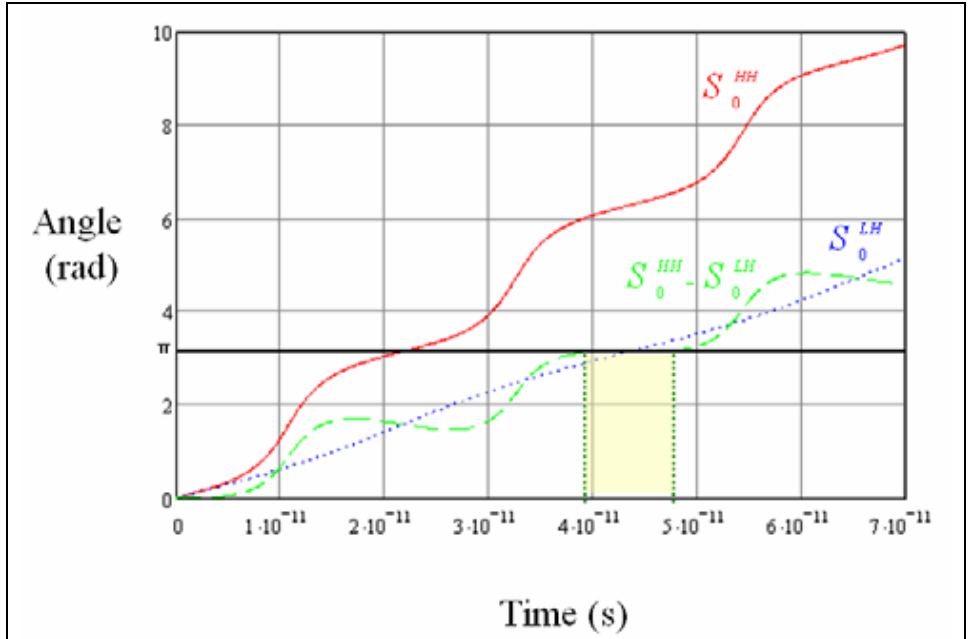

**Figure 15.**
Phases accumulated during interaction as a function of time.

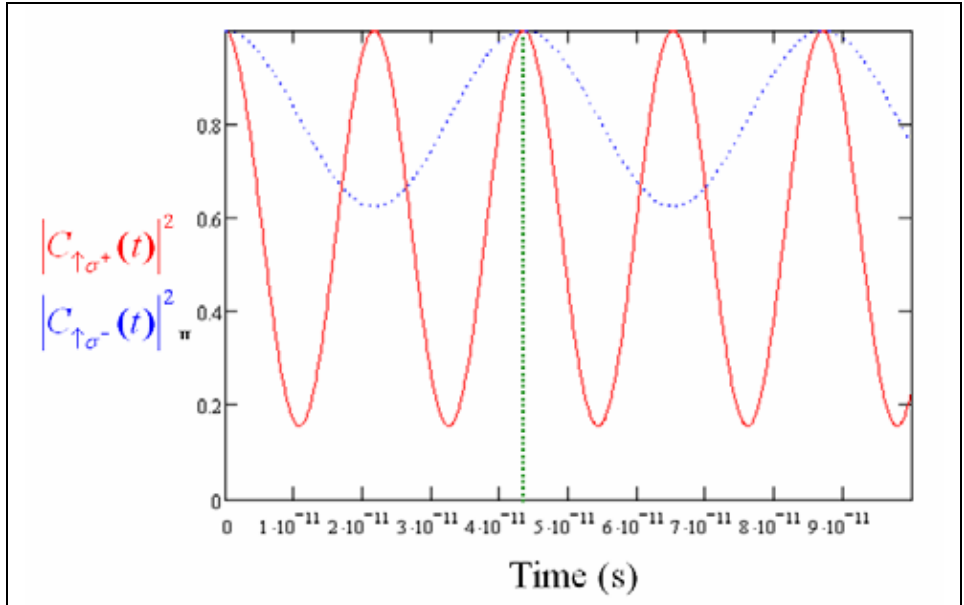

**Figure 16.**
Time evolution of the probability amplitudes

Interestingly, the phase requirement is satisfied over a range of interaction times, which is from approximately $T = 39\,ps$ to $T = 47\,ps$, while the probability amplitude requirements are satisfied exactly at $T = 43\,ps$ with a very small variation within $\pm 4\,ps$. This indicates that at this optimized operating point, this scheme is very robust against possible phase error within the considered range above. Furthermore, this interaction time is also much smaller than the limiting *spin decoherence time* of 20ms in semiconductor nanostructures (Kikkawa and Awschalom 1998).

Next, the coupling constants can be derived from the following equations

$$g_{3/2v,1/2c} = \frac{\mu_{hh} \cdot E}{\hbar} \quad g_{1/2v,1/2c} = \frac{\mu_{lh} \cdot E}{\hbar} \tag{14}$$

where $E = \sqrt{\dfrac{\hbar \omega_{ph}}{2 \cdot \varepsilon o \cdot \varepsilon r \cdot V}}$ and $\mu_{hh} = 3 \cdot \mu_{lh}$. This gives $g_{3/2v,1/2c} = 21\,GHz$ and $g_{1/2v,1/2c} = 7\,GHz$, which correspond to $\Omega_{3/2} = 46\,GHz$ and $\Omega_{1/2} = 23\,GHz$. The target cavity Qs and the cavity-waveguide coupling constant can also be determined. A commonly accepted condition for the strong coupling regime (Khitrova et al. 2006) is

$$2 \cdot \Omega_{3/2} \geq \frac{(\Gamma + \gamma)}{2} \tag{15}$$

where $\Gamma$ is the cavity decay rate and $\gamma$ is the dipole dephasing rate. Equations for the cavity decay rate and the dipole dephasing rate are

$$\Gamma = \frac{\omega_{ph}}{Q}$$
$$\gamma = \gamma_{dot} + \gamma_{enh} = \gamma_{dot} + F_p \cdot \gamma_o = \gamma_{dot} + \left(\frac{3Q\lambda^3}{4\pi^2 n^3 V}\right) \cdot \gamma_o \tag{16}$$

where $\gamma_{dot}$ is the sum of the QD non-radiative dephasing rate and the radiative decay rate outside the cavity ($\gamma_o$), and $\gamma_{enh}$ is the enhanced rate due to the Purcell Effect. Therefore, for InAs/GaAs material system, the minimum Q for which the condition in equation is satisfied is approximately

6,000 using $\gamma_{dot} = 22\ GHz$, $F_p = 441$, and $\gamma_o = 0.088\ GHz$ (Khitrova et al. 2006). Recalling that an optimized L1 cavity can easily reach Qs of 17,000, the strong coupling regime is therefore achievable. If we assume that the cavity is designed to have a Q of 12,000, well within the strong coupling regime, the associated cavity decay rate is $\Gamma = 21 GHz$. While this presents an ideal system for the SPFE, it represents a potential barrier for the release of the single photon back into the SMZ arm. Release of the single photon by the nanocavity requires a transition into the weak coupling regime by increasing $\Gamma$, thus necessitating a mechanism to control the Q of the cavity in order to be able to release the photon back into the waveguide. Several methods can be envisioned - the idea of using a Q-switch cavity is attractive. Recently, dynamic control of a photonic crystal cavity Q factor was demonstrated by switching the cavity Q from 12,000 to 3,000 in the picoseconds regime (Tanaka et al. 2007). This represents a much more attractive method - once the cavity Q is switched to 3,000, the process of spontaneous emission is not reversible (i.e. in the weak coupling regime), resulting in an escape probability of 92% after a time t = 30ps using **Equation 17**. This alone provides much support for the feasibility of such devices.

$$P(\%) = 1 - e^{-\frac{t}{\tau}} \quad \text{with} \quad \tau = \frac{1}{\Gamma} \qquad (17)$$

*4.2 Performance*

In order to estimate the *fidelity*, the error $\delta_\varphi$ in the relative phase $\varphi$ (see **Equation 9** and **Equation 10**) must be calculated. It consists of both the phase errors due to the switching ($\delta_{\varphi,Qswitch}$) and due to the escape time ($\delta_{\varphi,Escape}$), which are expressed in **Equation 18**. The phase error due to the switching can be obtained by calculating the ratio of the switching time over the interaction time. In case of the phase error due to the escape time, the ratio of the rate of rotation of the linear polarization over the cavity decay rate. The rate of rotation is approximately the slope $m_{S^{lh}}$ of the phase accumulated during the interaction with the light holes ($S_O^{LH}$) in **Figure 15**.

$$\delta_\varphi = \delta_{Qswitch} + \delta_{\varphi,Escape}$$
$$\delta_{Qswitch} = \frac{\Delta t_{Qswitch}}{T} \qquad (18)$$
$$\delta_{\varphi,Escape} \approx \frac{m_{S^{lh}}/2\pi}{\Gamma}$$

Here, given the interaction time $T = 43\ ps$, the time to switch the cavity Q from 12,000 to 3,000 $\Delta t_{Qswitch} = 4\ ps$ (Tanaka et al. 2007), the cavity decay rate $\Gamma = 84\ GHz$ at Q=3,000, and the rate of rotation of the linear polarization $\frac{m_{S^{lh}}}{2\pi} = 11.3\ GHz$, the following relative phase error is obtained $\delta_\varphi = 0.026 + 0.1345 = 0.1605$ or 16.5% ($\delta_{Qswitch}$ could really be ignored since the phase does not change much in a time range of $\pm 4\ ps$).

There exists a quick solution to reduce the total phase error; it is simply to reduce the cavity Q to 10,000. Since the dynamic range $\Delta Q = 9,000$ is achievable, the cavity Q at the moment of the switch will be 1,000. This corresponds to an escape probability of 92% after a time t = 10ps. The phase error is then $\delta_\varphi = 0.026 + 0.044 = 0.070$ or 7%. This method modifies only phase error due to the escape time.

Another solution is to increase the detuning energy to $E_d = 1.0 meV$ for instance. However, doing so will reduce the rate of rotation of the linear polarization. Since the detuning is much larger than the coupling rates ($\Delta \gg g_{3/2v,1/2c}, g_{1/2v,1/2c}$) between the photon and the QD, the interaction rates are determined using second order perturbation theory. The new coupling strengths can be expanded in the following manner

$$\Omega_{3/2} = \sqrt{\Delta^2 + 4 \cdot g_{3/2v,1/2c}^2} \approx \Delta + \frac{2 \cdot g_{3/2v,1/2c}^2}{\Delta} + \ldots$$

$$\Omega_{1/2} = \sqrt{\Delta^2 + 4 \cdot g_{1/2v,1/2c}^2} \approx \Delta + \frac{2 \cdot g_{1/2v,1/2c}^2}{\Delta} + \ldots$$
(19)

The expressions for the accumulated phases during interaction with the QD take the following forms

$$S_0^{HH} = \tan^{-1}\left(\frac{\Delta}{\Omega_{3/2}} \cdot \tan(\frac{\Omega_{3/2}t}{2})\right) \approx \tan^{-1}\left(1 \cdot \tan(\frac{\Delta + \frac{2 \cdot g_{3/2v,1/2c}^2}{\Delta}}{2} t)\right) = \left(\frac{\Delta}{2} + \frac{g_{3/2v,1/2c}^2}{\Delta}\right) t$$

$$S_0^{LH} = \tan^{-1}\left(\frac{\Delta}{\Omega_{1/2}} \cdot \tan(\frac{\Omega_{1/2}t}{2})\right) \approx \tan^{-1}\left(1 \cdot \tan(\frac{\Delta + \frac{2 \cdot g_{3/2v,1/2c}^2}{\Delta}}{2} t)\right) = \left(\frac{\Delta}{2} + \frac{g_{1/2v,1/2c}^2}{\Delta}\right) t$$
(20)

The terms $\frac{\Delta}{2}$ can be ignored because they do not contribute to the rotation of the linear polarization, since they cancel each other out when the two phases are subtracted. The expressions for the interaction frequencies are thus

$$\Omega_{3/2} = \frac{S_0^{HH}}{t} \cong \frac{g_{3/2v,1/2c}^2}{\Delta}$$

$$\Omega_{1/2} = \frac{S_0^{LH}}{t} \cong \frac{g_{1/2v,1/2c}^2}{\Delta}$$
(21)

Interestingly, the interaction frequencies are no longer required to be multiples of $\pi$, since the admixture to the exciton state is negligible. Additionally, the condition for the strong coupling regime in **Equation 15** must be re-derived for the new interaction frequencies.

With the modified interaction rates $\Omega_{3/2} = 11.5 MHz$ and $\Omega_{1/2} = 3.8 MHz$, the interaction time needed was calculated to be $T = 1.2 ns$ and a new minimum cavity Q about 11,000 is needed for the strong coupling to occur The enhanced rate due to the Purcell effect was ignored in calculating the new condition for the strong coupling regime due to the highly detuned photon energy (Ryu and Notomi 2003).The original L1 cavity with a Q of 12,000 resulting in a of Q of 3,000 when it is switched can still be used. The new phase error is $\delta_\varphi = 0.003 + 0.0145 = 0.0175$ or $1.75\%$ using **Equation 22**.

$$\delta_\varphi = \delta_{Qswitch} + \delta_{\varphi, Escape}$$

$$\delta_{Qswitch} = \frac{\Delta t_{Qswitch}}{T}$$

$$\delta_{\varphi, Escape} = \frac{\Omega_{3/2} + \Omega_{1/2}}{\Gamma}$$
(22)

This method affects both the phase error due to the switch and the one due to the escape time. The phase error can be further reduced at the expense of lower rates of rotation of the linear polarization, higher cavity Qs, and therefore higher dynamical Q-switching ranges.

Polarizers could in theory be used within each arm of the switch to eliminate the relative phase error $\delta_\varphi$. Using **Equation 9** and **Equation 10**, the new relative phase can be expressed in the following form

$$\varphi_{new} = \varphi + \varphi_{Error} = \varphi + (\delta_\varphi \cdot \varphi) \tag{23}$$

Ignoring the general phase term $\theta$, the X and Y components of the electric field vector can be rewritten including the phase error term $\phi_{Error} = \delta_\varphi \cdot \phi$ as shown below.

$$\vec{e} = \left(\cos(\varphi + \delta\phi)|x\rangle + \sin(\varphi + \delta\phi)|y\rangle\right) \tag{24}$$

Assuming $\delta_\varphi = 0.0175$ (see above) and $\varphi = \frac{\pi}{2}$ (the electric field is pointing in the Y direction upon the completion of the SPFE), then $\varphi + \delta\phi = 1.598\ rad$. If the Y-polarizer is utilized, the resulting attenuation due to the relative phase error is approximately 0.074% ($|\cos(\varphi + \delta\phi)|^2 = 0.000739$) of the intensity, which corresponds to an efficiency of 99.92%. This end result is significant and very promising for the realization of a quantum network based on this technology.

Furthermore, other losses are also traditionally small. Chen et al calculated propagation losses in photonic crystals and found a decaying constant $\beta = 0.05 cm^{-1}$ for the fundamental guided mode (propagating in the plane) in triangular lattice with air columns (Chen et al. 2005). This would correspond to an efficiency of 99.99% in term of propagation losses for small structures such as the SMZ interferometers (It is assumed that the probability for out-of-plane scattering is negligible). In addition, for the single qubit gates, highly efficient polarizing beam splitters can be used. For instance, Seunghyun Kim et al reported efficiencies above 99% for both TE and TM polarized light propagating through their photonic crystal based polarizing beam splitter (Kim et al. 2003).

Finally, the error $\delta_\theta$ in the general phase $\theta$ (see **Equation 9** and **Equation 10**) and its effect on both the switch during the ON and OFF stages as well as the single qubit gates is investigated. The general phase can be understood as an *effective path length difference* between the two arms of the SMZ interferometer and cannot be corrected using a polarizer. It contributes to the *insertion losses* or attenuation when the switch is ON; whereas, it is responsible for some *optical leakage* when the switch is OFF. Equation 25 below shows the expression for the general phase error.

$$\begin{aligned}
\delta_\theta &= \delta_{Qswitch} + \delta_{\theta,Escape} \\
\delta_{Qswitch} &= \frac{\Delta t_{Qswitch}}{T} \\
\delta_{\theta,Escape} &\approx \frac{(m_{S^{hh}} + m_{S^{lh}})/2\pi}{\Gamma} \qquad \left(\Delta \simeq g_{3/2v,1/2c}, g_{1/2v,1/2c}\right) \\
\delta_{\theta,Escape} &= \frac{\Omega_{3/2} + \Omega_{1/2}}{\Gamma} \qquad \left(\Delta \gg g_{3/2v,1/2c}, g_{1/2v,1/2c}\right)
\end{aligned} \tag{25}$$

If the cavity Q at the time of the switch is assumed to be as low as 1,000, the general phase error is $\delta_\theta = 0.026 + 0.15 = 0.176$ (17.6%) for a detuning energy $E_d = 75\mu eV$, or $\delta_\theta = 0.003 + 0.0096 = 0.0126$ (1.26%) for a detuning energy $E_d = 1meV$. Thus, if the switch is OFF and the detuning energy is $E_d = 1meV$, then the general phase with the error is $\theta = \pi + (\delta_\theta \cdot \pi) = 3.18\ rad$ which means that 0.039% of the intensity goes through ($\frac{1+\cos(\theta)}{2} = 0.000391$) corresponding to a 34-dB extinction ratio. On the other hand, if the switch is ON ($\theta = 0 + (\delta_\theta \cdot \pi) = 0.039 rad$), 0.039% of the intensity is lost on account of the error in the general phase, resulting in an efficiency of 99.96%. This corresponds to an insertion loss of only -0.0017-dB. This general phase error also causes the linear polarization of a photon qubit to become somewhat elliptical following a single qubit operation. This effect is small and does not greatly affect the fidelity.

## 5. Conclusion

We have introduced a new concept for a SMZ interferometer, based on the single photon Faraday Effect, which can play a key role as a component for future quantum information devices or quantum networks. We demonstrated how the SMZ interferometer could be used as a switch to regulate the flow of quantum information both within a quantum computer and within a quantum communication system using a Wavelength-Division-Multiplexing scheme. In addition, we introduced cases where SMZ interferometers could also be used as single quantum gates such as the X, Z, and H gates. In comparison with other proposed realizations of quantum computing (nuclear spin, harmonic oscillator, superconductors, etc) this approach ranks second in term of the maximum number of operations with $n_{op} = 10^8$ operations, which is defined as decoherence time ($10^{-3}$ seconds) over the interaction time ($10^{-12}$ seconds) (Kroutvar et al. 2004)

## Acknowledgements

We would like to acknowledge the support for this work from the National Science Foundation under grant number ECCS-0725514.